\documentstyle[12pt]{article}

\begin{document}

\begin{titlepage}
\rightline {Si-94-13 \  \  \  \   }

\vskip 3.truecm

\centerline{\large \bf  Dynamic Monte Carlo Measurement
 of  Critical Exponents }
\vskip 2.0truecm
\centerline{\bf Z.B. Li\footnote {On leave of absence
from Zhongshan University,
510275 Guangzhou, P.R. China},  L. Sch\"ulke, and B. Zheng }

\vskip 0.7truecm

\centerline{Universit\"at-GH Siegen, D-57068 Siegen, Germany}

\vskip 1.truecm

\centerline{\large October 27, 1994}

\vskip 2.7truecm

\abstract{
             Based on the scaling relation for the dynamics at the
      early time, a new method is proposed to measure both the
      static and dynamic critical exponents. The method is
      applied to the
      two dimensional Ising model. The results
      are in good agreement with
      the  existing results.
      Since the measurement is carried out in the initial
      stage of the relaxation process starting from
      independent initial
      configurations, our method is efficient.}


\end{titlepage}

Numerical measurements of critical exponents
are usually carried out from samples of
configurations in the equilibrium
generated with Monte Carlo algorithms.
For the static exponents,
Binder's method is one of the widely accepted
\cite{bin92,bin81}. A traditional way
to obtain the dynamic exponent $z$ is to measure
the exponential
decay of the time correlations for finite systems
in the long-time regime \cite{wan91,wil85}.
At the critical point,
except for some special algorithms as e.g. the cluster algorithm
\cite {swe87,wol89},
one suffers from critical slowing down. Recently, it has been
suggested that $z$ may be estimated from the
power law decay of the magnetization in
a relaxation process on a large enough lattice,
in a sufficient large time but before the exponential decay starts
\cite{mun93,sta92}.

In the past few years, better understanding has been
achieved on the critical relaxation processes
even upto the early time
of the evolution. A representative example of these processes
is that the
Ising model initially in random states with a small magnetization
is suddenly quenched to the critical temperature
and then evolves with the dynamics of model A.
It was shown by Janssen et al.
\cite {jan89} with $\varepsilon$-expansion
upto two-loop that
besides the well-known long-time universal
behaviour,
there exists another {\it universal} stage
of the relaxation at earlier time, termed
``{\it critical initial slip}'', which sets
in right after the microscopic time scale and
eventually crosses over to the long-time regime.
For the critical initial slip the characteristic time scale is
$t_{0}\sim m_{0}^{-z/x_{0}}$ with $x_{0}$ being a new independent critical
exponent and $ m_{0}$ the initial magnetization.
The scaling behaviour
including the increase of the order
has been illustrated by
a number of authors by Monte Carlo simulations\footnote{
After the present letter was completed,
we received a preprint of
P. Grassberger (Wuppertal, FRG)
where the critical initial slip
related to damage spreading was investigated and
the dynamic exponent $z$
also measured with the Monte Carlo method
in both two and three dimensional Ising model.}
\cite{bra91,hus89,men94,li94},
 and also
analytical calculations \cite{die93,jan93,jan94,rit94}.

Of special interest is here
the extension of the results in Ref. \cite{jan89} to
finite-size systems \cite{die93,li94}.
In accordance to the renormalization group analysis
for finite-size
systems,
we expect a scaling relation to hold for the k-th moment of the
magnetization in the neibourhood of the critical point
\cite{pri84,jan89,zin89},
\begin{equation}
M^{(k)}(t,\tau,L,m_{0})=b^{-k\beta/\nu}
M^{(k)}(b^{-z}t,b^{1/\nu}\tau,b^{-1}L,
b^{-x_{0}}m_{0})
\label{mkscal}
\end{equation}
where $t$ is the evolution time,
$\tau=(T-T_{c})/T_{c}$ is the reduced temperature,
$L$ is the lattice size,
and $b$ is the spatial rescaling factor. It has been stressed in
Ref. \cite{jan89} that the initial states must have very {\it short}
correlation lengths and the initial magnetization $m_{0}$ must be
{\it sharply} prepared.
The scaling relation for the
3-dimensional Ising model has been tested by Monte Carlo
simulation \cite{li94}. The numerical data fit into the scaling
relation
nicely, and $x_{0}$ has been determined with satisfactory precision.
It was observed that
the microscopic time scale is ignorably small, for example,
for the heat-bath Monte Carlo
algorithm it is smaller than one MC sweep.
This clean behaviour
of the critical relaxation in its {\it early time} indicates
a promising
new way to measure both the {\it static} and {\it dynamic}
critical exponents, which is similar to
Binder's method in equilibrium. In the present letter, we will
illustrate this idea
for the two dimensional Ising model.

To make the computation simpler and more efficient,
we set $m_{0}$ to its
fixed point, $m_{0}=0$. Therefore the exponent $x_{0}$
will not enter
the calculation. Furthermore, now the time scale
$t_{0}=m_0^{-z/x_0} \to \infty$, and the
critical initial slip gets most prominent in time direction
even though the magnetization itself will only fluctuate
around zero.
The choice of $m_{0}=0$ is essential in the calculation,
which allows
a more precise and stable measurement of the critical exponents.
After generating randomly an initial configuration
the system is released to the evolution
with the heat-bath
algorithm at the critical temperature.
We repeat this process with independent
initial configurations. The average is taken
over the initial configurations with $m_{0}=0$ and
zero correlation length.

To determine $z$, we introduce a {\it time-dependent}
 Binder cumulant
\cite{bin81}
\begin{equation}
U(t,\tau,L)\equiv 1-{M^{(4)} \over 3(M^{(2)})^{2}}.
\label{cumul}
\end{equation}
Following the scaling relation in Eq.(\ref{mkscal}),
cumulants measured in two different lattices
have the simple relation
\begin{equation}
U(t,0,L_{1})=U(b^{-z}t,0,L_{2})
\label{urel}
\end{equation}
with $b=L_{1}/L_{2}$. The exponent $z$ can easily be obtained through
searching for a time rescaling factor $b^{-z}$ such that the two curves
represented by both sides in Eq.(\ref{urel}) collapse.
In other words, the cumulant can be described by a scaling
function
\begin{equation}
f(t/L^{z})=U(t,0,L).
\label{uscal}
\end{equation}

\begin{figure}[t]\centering
\caption{
The cumulants $U_L$ for $L=8$, $16$ and $32$ (labelled by $+$, $\diamond$
and $\star$, respectively) together with the rescaled cumulants of
double lattice (solid lines).
}
\label{F1}
\end{figure}

In the same way,
it is easy to obtain other two scaling functions
\begin{equation}
g^{(2)}(t/L^{z})=L^{2\beta/\nu}M^{(2)}(t,0,L),
\label{betscal}
\end{equation}
\begin{equation}
h(t/L^{z})=L^{-1/\nu}\partial_{\tau}
\log M^{(2)}(t,\tau,L)\vert_{\tau=0}.
\label{nuscal}
\end{equation}
With $z$ in hand, $\beta/\nu $ and $\nu$ can be determined by
fitting respectively the scaling function
$g^{(2)}(t/L^{z})$ and $h(t/L^{z})$
from different lattices similar as in
 the determination of $z$.
We adopt the derivative of $M^{(2)}(t,\tau,L)$  for the estimation
of $\nu$, rather than that of $U(t,\tau,L)$
since the measurement of $M^{(2)}$ is more stable
for large $L$.
To our knowledge, so far the measurment of the static and dynamic
exponents from the dynamic process in the early time
has not been investigated.
One can easily realize that at time $t \to \infty$
Binder's way to calculate $\beta/\nu$ and $\nu$
is recovered\footnote{With a similar procedure used by Binder,
the critical point can also be located from the behaviour
of $U(t,\tau,L)$ around $\tau =0$.} \cite {bin92}.
Since the measurement can now be carried out already in the initial stage
of the relaxation, the method becomes more efficient.

In Fig.~1, we plot the cumulants of the two-dimensional Ising model
versus time. The curves  corresponding to
$L=8,16$ and $32$, repectively, are labelled by
$+$,$\diamond$ and $*$. Each curve is compared with
the time-rescaled
cumulant of double lattice size given by solid lines.
The exponents corresponding to the three
pairs of best-fitted curves are $z=2.0969,\ 2.1493$,
and $2.1337$, respectively.
Our best value $2.1337$ should be compared with the existing
numerical results from \cite {wil85}, $z=2.13(8)$
and from \cite {lan87}, $z=2.14(5)$ and also with that with the
$\epsilon$-expansion from \cite {bau81}, $z=2.126$.
{}From the figure one can also see clearly that the scaling
relation holds  from the very beginning of the time evolution and
how remarkably well the method is working.

In Tab.~1, we have summarize the measured exponents $z$,
$2\beta/\nu$ and $1/\nu$ for
the 2-dimensional Ising model. They are
in good agreement with
the exact results \cite{bax82} and best values for $z$
\cite{wil85,bau81,lan87}.
To obtain these results, we have
averaged over $50\;000$
independent samples in each run  and repeated $8$ runs
to estimate the errors.

\begin{table}[t]\centering
$$
\begin{array}{|c|l|l|l|}
\hline
L_1\leftrightarrow L_2  &\qquad  z & \quad2\beta/\nu &\quad 1/\nu \\
\hline
\ 8\leftrightarrow16 & 2.0969(20) & 0.2480(02) &  1.127(08)  \\
16\leftrightarrow32 & 2.1493(18) & 0.2494(06) &   1.058(35)  \\
32\leftrightarrow64 & 2.1337(41) & 0.2504(29)&   0.955(40)  \\
\hline
\end{array}
$$
\caption{
 Results for $z$, $2\beta/\nu$ and $1/\nu$, respectively,
from the $2$-dimensional Ising model.
}
\label{T1}
\end{table}

Compared with traditional measurements
in equilibrium,  much less effort is needed with
our $dynamic$ Monte Carlo algorithm, since we do not enter
the long-time regime where critical slowing down is severe.
The efficiency of our method may also be traced back
to the simple power law increase of the moments
 at the early time.
For example, it is well known that $M^{(2)}(t,0,L) \sim
t^{(d-2\beta/\nu)/ z }$ when $L$ is large enough
\cite {hus89,bra91}. Therefore the scaling relation
implies $M^{(2)}(t,0,L) \sim
L^{-d}\, t^{(d-2\beta/\nu)/z }$.
 Similar arguments exists for the cumulant
$U(t,\tau,L)$.

In  previous works on the critical initial slip
\cite{jan89,die93,li94,rit94}, the increase of the order
and the role of $m_0$ in the short- and long-time regimes
are intensively discussed. We would like, however, to
stress that
even in the case of $m_0=0$ it is interesting
and there exist fruitful applications,
as reported in this letter.
Our results also provide
a further confirmation of the scaling relation discovered
by Janssen et al. \cite{jan89}. Investigations for dynamics
in other universality classes should be
carried out.

\vskip 1. truecm

{\it Acknowledgement:} The authors would like to thank
   U. Ritschel for handing useful preprints and
   helpful discussions in the previous
   collaboration. Z.B.L. is grateful to the Alexander von
Humboldt-Stiftung for a fellowship.

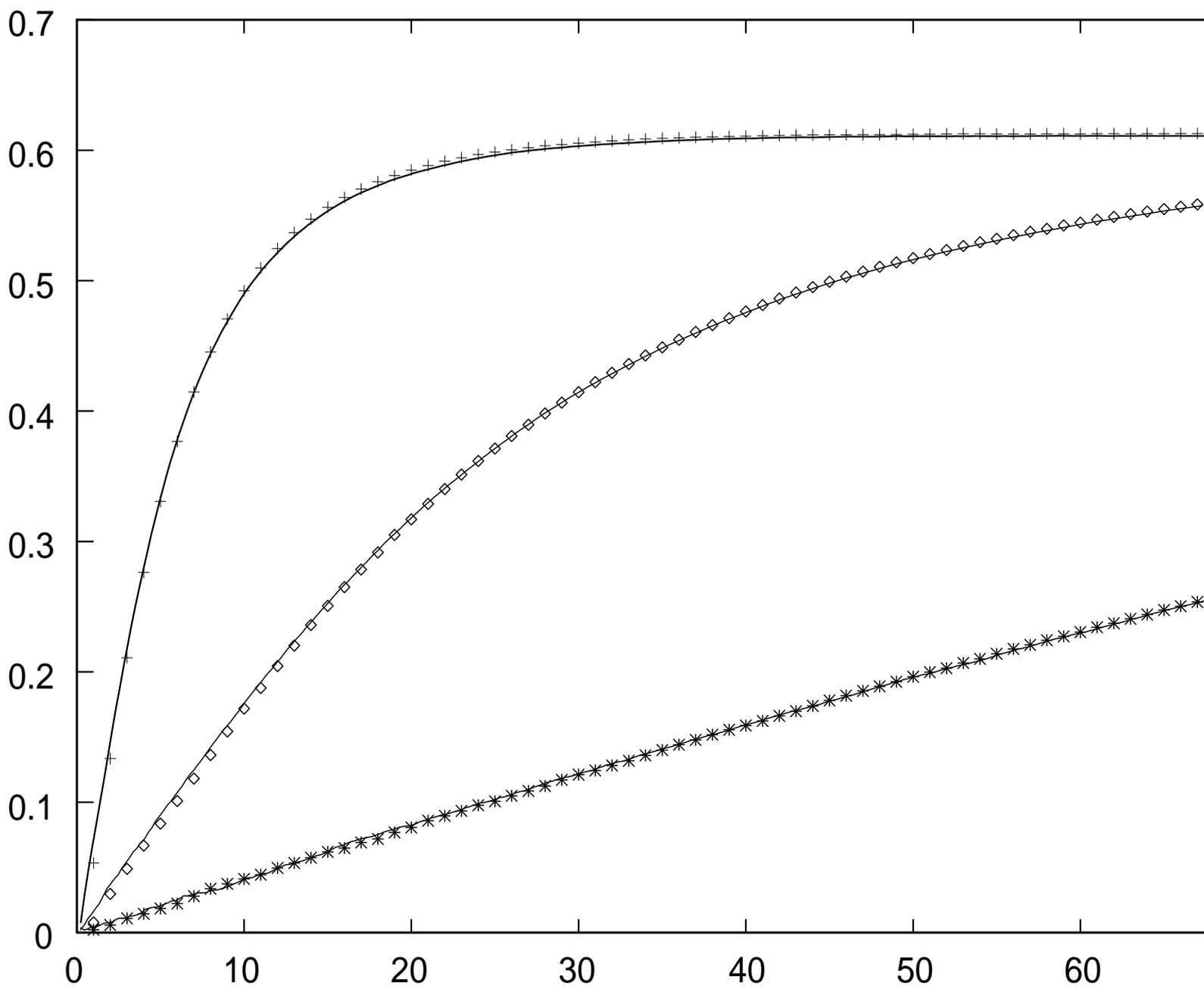
\end{document}